\documentclass{aa}
\usepackage{txfonts}
\usepackage{graphicx}
\usepackage{natbib}
\begin{document}

\title{Probabilities of exoplanet signals from posterior samplings}
\author{Mikko Tuomi\inst{1,2}\thanks{The corresponding author, \email{m.tuomi@herts.ac.uk; mikko.tuomi@utu.fi}} \and Hugh R. A. Jones\inst{1}}
\institute{University of Hertfordshire, Centre for Astrophysics Research, Science and Technology Research Institute, College Lane, AL10 9AB, Hatfield, UK \and University of Turku, Tuorla Observatory, Department of Physics and Astronomy, V\"ais\"al\"antie 20, FI-21500, Piikki\"o, Finland}
\date{Received XX.XX.2011 / Accepted XX.XX.XXXX}

\abstract{}
{Estimating the marginal likelihoods is an essential feature of model selection in the Bayesian context. It is especially crucial to have good estimates when assessing the number of planets orbiting stars and the different models explain the noisy data with different numbers of Keplerian signals. We introduce a simple method for approximating the marginal likelihoods in practice when a statistically representative sample from the parameter posterior density is available.}
{We use our truncated posterior mixture estimate to receive accurate model probabilities for models with differing number of Keplerian signals in radial velocity data. We test this estimate in simple scenarios to assess its accuracy and rate of convergence in practice when the corresponding estimates calculated using deviance information criterion can be applied to receive trustworthy results for reliable comparison. As a test case, we determine the posterior probability of a planet orbiting HD 3651 given Lick and Keck radial velocity data.}
{The posterior mixture estimate appears to be a simple and an accurate way of calculating marginal integrals from posterior samples. We show, that it can be used to estimate the marginal integrals reliably in practice, given a suitable selection of parameter $\lambda$, that controls its accuracy and convergence rate. It is also more accurate than the one block Metropolis-Hastings estimate and can be used in any application because it is not based on assumptions on the nature of the posterior density nor the amount of data or parameters in the statistical model.}
{}

\keywords{Methods: Statistical, Numerical -- Techniques: Radial velocities -- Stars: Individual: HD 3651}

\titlerunning{Probabilities of exoplanet signals}

\maketitle


\section{Introduction}

The selection between a collection of candidate models is of significant in all fields of astronomy but especially so, when the purpose is to extract weak planetary signals from noisy data. The ability to tell whether a signal is present in data as reliably as possible is essential in several searches for low-mass exoplanets orbiting nearby stars, whether made using the Doppler spectroscopy method; e.g. the Anglo-Australian Planet Search \citep[e.g.][and references therein]{tinney2001,jones2002}, High-Accuracy Radial Velocity Planet Searcher \citep[e.g.][and references therein]{mayor2003,lovis2011}, Hich Resolution Echelle Spectrometer \citep[e.g.][and references therein]{vogt1994,vogt2010}; by searching photometric transits; e.g. Convection Rotation and Planetary Transits \citep[e.g.][and references therein]{barge2007,hebrard2011}, WASP \citep[e.g.][and references threrein]{colliercameron2007,faedi2011}; or other possible techniques, such as astrometry \citep[e.g.][]{benedict2002,pravdo2009} and transit timing \citep[e.g.][]{holman2005} or other current or future methods.

Using Bayesian tools, it is possible to determine the relative probabilities for each statistical model in some selected collection of models to assess their relative performance, or relative ability to explain the data in a probabilistic manner. This is also important in the context of being able to assess their inability to explain several data sets in terms of the model inadequacy of \citet{tuomi2011b}. Especially, when different statistical models contain different numbers of planets orbiting the target star, assessing their relative posterior probabilities given the measurements is extremely important to detect all the signals in the data \citep[e.g.][]{gregory2005,gregory2007a,gregory2007b,tuomi2009,tuomi2012} and to avoid the detection of false positives \citep[e.g.][]{bean2010,tuomi2011}. However, determining the posterior probabilities require the ability to calculate marginal integrals that are complicated multidimensional integrals of likelihood functions and priors over the whole parameter space. While there are several methods of estimating the values of these integrals, those that are computationally simple and easy to implement are more often than not the poorest ones with respect to their accuracy and convergence properties \citep[e.g.][]{kass1995,clyde2007,ford2007}. There are also more complicated methods for estimating multidimensional integrals but they may provide more difficult computational problems themselves than typical data analyses are, which makes it difficult to use them in practice.

Because of these difficulties and the need to be able to assess the marginal integrals reliably, we introduce a simple method for estimating the marginal integrals in practice if a statistically representative sample from the parameter posterior density exists. As such a sample is usually calculated when assessing the posterior densities of model parameters using posterior sampling algorithms \citep[e.g.][]{metropolis1953,hastings1970,haario2001}, the ability to use the very same sample in determining the marginal integral is extremely useful in practice. There are methods for taking advantage of the posterior sample in this manner \citep[e.g.][]{newton1994,kass1995,chib2001,clyde2007} but their performance, despite some studies \citep[e.g.][]{kass1995,ford2007}, is not generally well known, especially so in astronomical problems, and some of them may also require samplings from other densities simultaneously, such as the prior density or the proposal density of the Metropolis-Hastings (M-H) output, making their application difficult.

In this article, we introduce a simple method that can be used to receive accurate estimates for the marginal integral. We test our estimate, called the truncated posterior mixture (TPM) estimate in scenarios where the marginal integral can be calculated accurately using simple existing methods. The deviance information criterion \citep[DIC,][]{spiegelhalter2002} is asymptotically an accurate estimate when the sample size, i.e. the sample drawn from the posterior density, increases and can be used if the posterior is a multivariate Gaussian. Therefore, we compare our estimate with the DIC estimate in such cases to test its accuracy in practice. If accurate, our estimate is applicable whenever a statistically representative sample from the posterior is available because we do not make any assumptions regarding the shape of the posterior density when deriving the TPM estimate. The only assumptions are, that such a sample exists and it is statistically representative. We also calculate the marginal likelihoods using the simple Akaike information criterion (AIC) for small sample size \citep{akaike1973,burnham2002}, the harmonic mean (HM) estimate that is a special case of the TPM with poor convergence properties, and the One Block Metropolis-Hastings (OBMH) method of \citet{chib2001} that requires the simultaneous sampling of posterior and proposal densities. \citet{kass1995} and \citet{clyde2007} give detailed summaries of different methods in the context of model selection problems.

Finally, we also test the performance of the TPM estimate and the effects of prior choice in simple cases where it is possible to calculate the marginal integral from a sample from the prior (with the common mean estimate) and/or using direct numerical integration. Especially, we show the undesirable effects of Bartlett's paradox on the marginal integrals and demonstrate that the TPM estimate actually circumvents these effects in practice.

\section{Estimating marginal integrals}

In the Bayesian context, the models in some \emph{a priori} selected collection can be equipped with relative numbers representing the probabilities of having observed the data $m$ if the model was a correct one. Therefore, for $k$ different models $\mathcal{M}_{1}, ..., \mathcal{M}_{k}$, these probabilities are calculated as
\begin{equation}\label{model_probability}
  P(\mathcal{M}_{i} | m) = \frac{P(m | \mathcal{M}_{i})P(\mathcal{M}_{i})}{\sum_{j=1}^{k} P(m | \mathcal{M}_{j})P(\mathcal{M}_{j})} ,
\end{equation}
where $P(\mathcal{M}_{i})$ are the prior probabilities of the different models and the marginal integrals, sometimes called the marginal likelihoods, are defined as
\begin{equation}\label{marginal}
  P(m | \mathcal{M}_{i}) = \int l(m | \theta_{i}, \mathcal{M}_{i}) \pi(\theta_{i} | \mathcal{M}_{i}) d \theta_{i}
\end{equation}
and $l$ denotes the likelihood function and $\pi(\theta | \mathcal{M}_{i})$ is the prior density of the parameters.

The truncated posterior mixture estimate that approximates the marginal integral is defined as (see Appendix)
\begin{eqnarray}\label{TPM_estimate_text}
 && \hat{P}_{TPM} = \Bigg[ \sum_{i=1}^{N} \frac{l_{i}p_{i}}{(1-\lambda) l_{i}p_{i} + \lambda l_{i-h}p_{i-h}} \Bigg] \nonumber\\
 && \times \Bigg[ \sum_{i=1}^{N} \frac{p_{i}}{(1-\lambda) l_{i}p_{i} + \lambda l_{i-h}p_{i-h}} \Bigg]^{-1} .
\end{eqnarray}
where $l_{i}$ is the value of the likelihood function at $\theta_{i}$, $p_{i}$ is the value of prior density at $\theta_{i}$, and $\lambda \in [0,1]$ and $h \in \mathbb{N}$ are parameters that control the convergence and accuracy properties of the estimate. While it is easy to select $h$ -- it only needs to be large enough such that $\theta_{i}$ and $\theta_{i-h}$ are independent -- selecting parameter $\lambda$ is more difficult. If $\lambda$ is too large, the sample from the posterior is not close to the sample from the importance sampling function $g$ in the Eq. (\ref{definition_g}) in the Appendix, and the resulting estimate for the marginal is biased. Conversely, too small values of $\lambda$, while making the estimate more accurate, decrease its convergence rate because the estimate approaches asymptotically the HM estimate that is known to have extremely poor convergence properties \citep[see the Appendix and][]{kass1995}. Therefore, we test different values of $\lambda$ to find the best choice in applications. We note, however, that when $\theta_{i}$ and $\theta_{i-h}$ are independent, i.e. when $h$ is large enough given the mixing properties of the Markov chain used to draw a sample from the posterior density, the TPM can converge to the marginal integral. The reason is that it is clear from Eq. (\ref{TPM_estimate_text}) that occasional very small values of $l_{i}$, that consequently have a large impact on the sums in the estimate, do not slow down the convergence as much as they would in the HM estimate because it is unlikely that $l_{i-h}$ is also small at the same time. This is the key feature in the TPM estimate that ensures its relatively rapid convergence in practice. 

We estimate the integral in Eq. (\ref{marginal}) using five methods. The HM estimate (see Appendix), the truncated posterior mixture estimate introduced here, the DIC, AIC, and the OBMH method of \citet{chib2001}. While the DIC is a reasonably practical estimate in certain cases, it requires that the posterior is unimodal and symmetric and can be approximated as a multivariate Gaussian density, which is only rarely the case in applications. It can be easily calculated by using the average of the likelihoods and the likelihood of the parameter mean, which also reveals why the posterior needs to be unimodal and symmetric for reliable results. These means do not reflect the properties of the posterior in the cases of skewness and multiple modes, not to mention nonlinear correlations between some parameters in vector $\theta$. The DIC is asymptotically accurate when the sample size becomes large \citep{spiegelhalter2002}. We do not consider the HM estimate to be a trustworthy one but calculate its value because it is a special case of the truncated posterior mixture estimate when $\lambda = 0$ (or 1). The AIC could provide a reasonably accurate estimate in practice, and therefore we compare its performance in various scenarios. However, it relies on the maximum likelihood parameter estimate, and does not therefore take into account the prior information on the model parameters. Its accuracy also decreases as the amount of parameters in the model increases or the number of measurements decreases. Finally, we calculate the OBMH estimate \citep{chib2001}. While this estimate appears to provide reliable results, e.g. the number of companions orbiting Gliese 581 determined in \citet{tuomi2011} was supported by additional data \citep{forveille2011}, its performance has not been studied throughly with examples. It is also computationally more expensive than the TPM estimate, and indeed the other estimates compared here, because it requires the simultaneous sampling from the proposal density of the M-H algorithm.

When assessing the convergence of our TPM estimate given some selection of $\lambda$, we say that it has converged if the estimate at the $i$th member of the Markov chain, namely $\hat{P}_{TPM}(i)$, satisfies $| \hat{P}_{TPM}(i+k) - \hat{P}_{TPM}(i) | < r$ for all $k > 0$ and some small number $r$ -- in accordance of the standard definition of convergence. However, in practice, we use the logarithms of $\hat{P}_{TPM}$ and a value of $r = 0.1$ on the logarithmic scale for simplicity. We also approximate the estimate as having converged if the convergence condition holds for $0 < k < 10^{5}$ for practical reasons. While all the estimates except the AIC (which is based only on the maximum likelihood value) converge the better the greater sample they are based on, we only plot this convergence for the TPM estimate. For DIC, HM, and OBMH, we calculate the final estimate using the mean and standard deviation of values from several samplings.

\section{Prior effects on marginal integrals}

Because the marginal integrals in Eq. (\ref{marginal}) are integrals over the product of likelihood function and prior probability density of the model parameters, the choice of prior has an effect on these integrals of different models. One such choice for standard model of radial velocity data was proposed in \citet{ford2007} and applied in e.g. \citet{feroz2011} and \citet{gregory2011}. Specifically, this prior limits the parameter space of jitter amplitude $\sigma_{j}$ to [0, $K_{0}$], that of reference velocity $\gamma$ to [-$K_{0}$, $K_{0}$], and that of velocity amplitude of the $i$th planet, $K_{i}$, to [0, $K_{0} (P_{min}/P_{i})^{1/3}$], where $P_{min}$ is the shortest allowed periodicity and $P_{i}$ is the orbital period of the $i$th planet. \citet{ford2007} propose that the hyperparameter $K_{0}$ should be set to 2129 ms$^{-1}$, which corresponds to a maximum planet-star mass-ratio of 0.01.

We assume for simplicity that $P_{min} = P_{i}$, which leads to a constant prior for the parameter $K_{i}$. It then follows that the prior probability density of a $k$-Keplerian model has a multiplicative constant coefficient proportional to $K_{0}^{-2-k}$ -- this corresponds to the hypervolume of the parameter space of the $k$-Keplerian model. Because this constant also scales the marginal integral in Eq. (\ref{marginal}), it can be seen that increasing $K_{0}$ can make the posterior probability of any planetary signal insufficient to claim a detection, because the ratio $P(m | \mathcal{M}_{k})/P(m | \mathcal{M}_{k-1})$ is proportional to $K_{0}^{-1}$.

The above can also be described in more general terms. In fact, as noted by \citet{bartlett1957} and \citet{jeffreys1961}, choosing a prior for any model with parameter $\theta$ such that $\pi(\theta) = c h(\theta)$ for all $\theta \in \Theta$, where $\Theta$ is the corresponding parameter space, can lead to undesired features with respect to model comparison results. Assume that this choice is made for model $\mathcal{M}_{1}$ but for a simpler model $\mathcal{M}_{0}$, for which parameter $\theta$ does not exist (the ``null hypothesis''), this prior does not exist either because the corresponding parameter is not a free parameter of the model. Then, the posterior probability of model $\mathcal{M}_{1}$ becomes
\begin{eqnarray}\label{bartlett_comparison}
  && P(\mathcal{M}_{1} | m) \propto P(m | \mathcal{M}_{1}) P(\mathcal{M}_{1}) \nonumber\\
  && = c P(\mathcal{M}_{1}) \int_{\theta \in \Theta} l(m | \theta, \mathcal{M}_{1}) h(\theta) d \theta , \nonumber\\
  && \textrm{where } c = \bigg[ \int_{\Theta} h(\theta) d \theta \bigg]^{-1} .
\end{eqnarray} 

Setting the prior constant such that $h(\theta) = 1$, yields $c = V(\Theta)^{-1}$, where $V(\Theta)$ denotes the hypervolume of the parameter space, and leads to the inconvenient conclusion that as the hypervolume of the parameter space $\Theta$ increases, the posterior probability of the model $\mathcal{M}_{1}$ decreases below that of the $\mathcal{M}_{0}$, which prevents the rejection of the null hypothesis regardless of the observed data $m$. This is called the Bartlett's paradox \citep{bartlett1957,kass1995} but it does not mean that improper and/or constant priors are useless nor that they should not be used in applications.

A convenient way around this ``paradox'', can be received by considering the definition of the parameters. Because the analysis results should not depend on the unit system of choice, nor the selected parameterisation, i.e. whether we choose parameter $\theta$ or $\theta' = f(\theta)$, where $f$ is an invertible (bijective) function, it is possible to choose the parameter system in a convenient way that makes $c = 1$ by transforming $\theta' = f(\theta)$ with some suitable $f$. For some choises of $f$ the constant prior of parameter $\theta$ does not correspond to a constant prior for $\theta'$, but we do not consider this well-known effect of prior choice further here.

For instance, if we apply this to a Gaussian likelihood with mean $g(\theta)$ (e.g. a superposition of $k$ Keplerian signals in radial velocity data) and variance $\sigma^{2}$, it becomes one with mean $g(f(\theta))$ and variance $\sigma^{2}$ -- this does not chance the posterior density of the parameters as we can always make the transformation back using $f^{-1}$ but a convenient choice of $f$ sets $c = 1$ and prevents the prior probability density of the parameters from having undesirable effects on the marginal integrals. A similar transformation of $\sigma$ is also possible as long as the $f(\sigma)$ retains the same units as the measurements have. Therefore, we are free to define the model parameters in any convenient way, using e.g. any unit system, and this, as long as we retain the same functional form in our statistical model, cannot be allowed to have an effect on the results of our analyses. Specifically, when analysing radial velocity data, choosing the unit system such that $K' = K K_{0}^{-1}$ does not change the posterior density nor the values the likelihood function has but it makes $\pi(K') = 1$ for all $K' \in [0,1]$, which does not result in different weights for the models with different numbers of planets. We demonstrate these effects further in Section 5 by analysing artificial data sets.

We note, that this procedure does not interfere with the Occam's razor that is a built-in feature of Bayesian analysis methods. It still holds, that as the number of free parameters in the statistical model increases, this model also becomes penalised the more heavily. The reason is, that increasing the dimension of the parameter space effectively increases the hypervolume that has a reasonably high posterior probability (but lower than the MAP estimate) given the data -- this increases the amount of low likelihoods in the posterior sample and in Eq. (\ref{TPM_estimate_text}), which in turn decreases the estimated marginal integral as it should in accordance with the Occamian principle of parsimony.

\section{Comparison of estimates: radial velocities of HD 3651}

To assess the performance of the TPM estimate for the marginal integral, we compare its performance with different selections of parameter $\lambda$ in simple cases where the marginal integral can be calculated reliably using the DIC, i.e. when the model parameters receive close-Gaussian posteriors and the sample size is large. Therefore, as test cases, we choose radial velocity time-series made using several telescope-instrument combinations that have different velocity offsets and different noise levels. The simple model without any Keplerian signals provides a suitable scenario where the DIC is known to be accurate and the accuracy of our estimate can be assessed in practice.

The nearby K0 V dwarf HD 3651 has been reported to be a host to a 0.20 M$_{\rm{Jup}}$ exoplanet with an orbital period of 62.23 $\pm$ 0.03 days and an orbital eccentricity of 0.63 $\pm$ 0.04 \citep{fischer2003}. The radial velocity variations of HD 3651 have been observed using the HIRES at the Keck I telescope \citep{fischer2003,butler2006} and the Shane and CAT telescopes at the Lick observatory \citep{fischer2003,butler2006}. These datasets contain measurements at 42 and 121 epochs, respectively. The reason we chose these data is that they enable us to investigate several scenarios reliably. The fact that the planet orbiting HD 3651 is on an eccentric orbit and there is plenty of data available make it possible to assess the accuracy of the TPM estimate in several scenarios by enabling the comparison to the DIC estimate that is accurate as long as the posterior density is Gaussian. Therefore, we investigate the accuracy and convergence properties of of the TPM in various scenarios: with high and low numbers of data compared to the number of model parameters, and when the marginal likelihoods of two models are close to each other and as far from each other as possible given the available data.

We analyse the radial velocities of HD 3651 made using the HIRES and Lick exoplanet surveys and calculate the marginal likelihoods of models with 0 and 1 Keplerian signals using the methods based on DIC, AIC, TPM, HM, and OBMH. We denote these estimates of the integral in Eq. (\ref{marginal}) as $\hat{P}_{DIC}$, $\hat{P}_{AIC}$, $\hat{P}_{TPM}$, $\hat{P}_{HM}$, and $\hat{P}_{OBMH}$, respectively. We also calculate the marginal integrals for a very simple case of 0-Keplerian model and HIRES data using a sample from the prior ($\hat{P}_{M}$) density and direct numerical integration ($\hat{P}_{D}$).

\subsection{Case 1. HIRES data}

\begin{figure}
\center
\includegraphics[angle=270, width=0.49\textwidth]{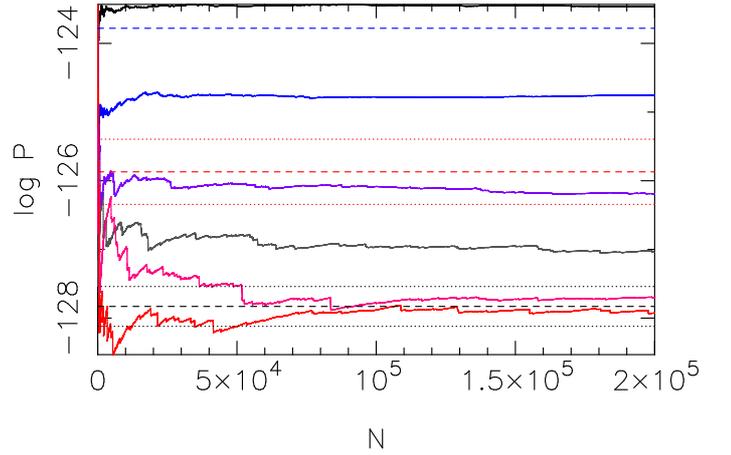}
\caption{Marginal integrals of the 1-planet model given the HIRES data (case 1): DIC and its 3$\sigma$ uncertainty (black dashed line and black dotted lines), AIC (blue dashed), OBMH and its 3$\sigma$ uncertainty (red dashed and red dotted), and the TPM estimates with $\lambda = 0.5, 0.1, 10^{-2}, 10^{-3}, 10^{-4}, 10^{-5}$ (black, grey, blue, purple, pink, and red curves).}\label{HIRES_marginals}
\end{figure}

The HIRES data with 42 epochs reveals some interesting differences between the five estimates for marginal integrals. The log-marginal integrals are plotted in Fig. \ref{HIRES_marginals} as a function of Markov chain length. The estimated uncertainties of DIC and OBMH estimates represent the standard deviations of six different Markov chains. The DIC estimate can be considered a reliable one in this case, because the posterior density is very close to a multivariate Gaussian. It can be seen that the AIC is biased because of the low number of measurements (42) compared to the number of parameters of the statistical model (7). Also, the OBMH estimate gives the 1-planet model a greater marginal likelihood than DIC. However, the TPM is similarly biased for $\lambda = 0.5, 0.1, 10^{-2}, 10^{-3}$ but converges to the DIC estimate for $\lambda = 10^{-4}, 10^{-5}$. The HM estimate is not shown in the Fig. because its extremely poor convergence properties -- it receives values between -130 and -140 on the logarithmic scale of Fig. \ref{HIRES_marginals}.

When using as small values of $\lambda$ for the TPM as possible such that it converges in the sense that it approaches some limiting value, we calculate the Bayes factors ($B$) in favour of the one-Keplerian model against the model without Keplerian signals. These values are shown in Table \ref{HIRES_bf}. The TPM estimate converges to the same value as DIC, which is known accurate in this case because the posterior densities of both models are very close to Gaussian. However, the AIC and OBHM overestimate the posterior probability of the model containing a Keplerian signal. Also, the problems of the HM estimate are clear because its uncertainty becomes greater than the estimated value.

\begin{table}
\center
\caption{Bayes factors in favour of the one-Keplerian model given the HIRES data (case 1).}\label{HIRES_bf}
\begin{tabular}{lc}
\hline \hline
Estimate & $B$ \\
\hline
TPM & 1.1$\times10^{14} \pm 1.2\times10^{13}$ \\
DIC & 1.1$\times10^{14} \pm 1.1\times10^{13}$ \\
AIC & 3.3$\times10^{15}$ \\
OBMH & 2.8$\times10^{16} \pm 5.6\times10^{15}$ \\
HM & 3.3$\times10^{13} \pm 6.5\times10^{13}$ \\
\hline \hline
\end{tabular}
\end{table}

\subsection{Case 2. Combined HIRES and Lick data}

Increasing the number of measurements likely makes the AIC yield a more accurate estimate for the marginal likelihood. However, to see how this affects the other estimates, we again compare them to the DIC which is reliable because of the close-Gaussianity of the posterior density. The inclusion of additional Lick data also makes the posterior probability of the one-Keplerian model much greater than that of the model without Keplerian signals, and enables us to investigate the accuracy and convergence of the TMP in such a scenario. Therefore, we study the properties of the different estimates for marginal integrals using the combined HIRES and Lick data of HD 3651 with 163 epochs.

\begin{figure}
\center
\includegraphics[angle=270, width=0.49\textwidth]{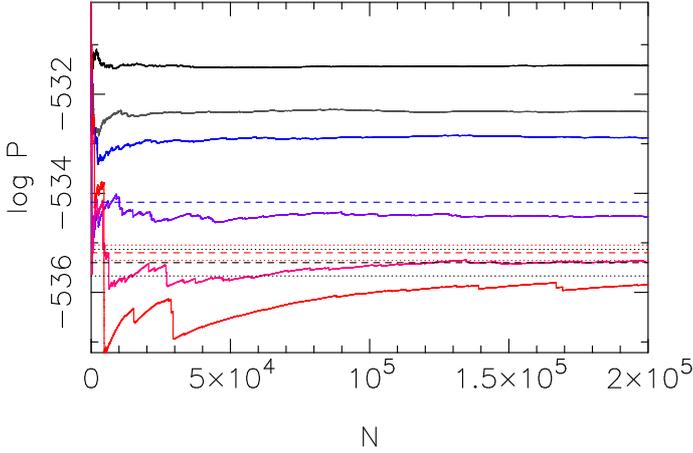}
\caption{As in Fig. \ref{HIRES_marginals} but for the combined data (case 2) and the TPM estimates with $\lambda = 0.1, 10^{-2}, 10^{-3}, 10^{-4}, 10^{-5}, 10^{-6}$ (black, grey, blue, purple, pink, and red curves).}\label{combined_marginals}
\end{figure}

TPM converges to the DIC estimate when $\lambda = 10^{-3}$ for the model without any Keplerians, whereas its convergence takes place for $\lambda = 10^{-5}$ for the one-Keplerian model (Fig. \ref{combined_marginals}, pink curve). Clearly, the AIC is indeed closer to the DIC estimate because of the greater number of data but the OBHM is also consistent with the DIC estimate. We note that the HM estimate is again omitted from the Fig. \ref{combined_marginals} because it receives significantly lower values than the other estimates.

Now, we calculate the Bayes factors in favour of the one-Keplerian model and present them in Table \ref{combined_bf}. The TPM estimate is again very close to the DIC estimate and the AIC is close to these, providing slightly greater support for the one-Keplerian model. The OBMH again overestimates the one-Keplerian model and the HM estimate, while this time being rather accurate, has an uncertainty in excess of the estimate itself. Clearly, the TPM estimate can be used to receive reliable estimates for the marginal integral in this case as well, because the posterior density is again very close to Gaussian and the DIC estimate is therefore a reliable one in assessing the integral.

\begin{table}
\center
\caption{Bayes factors in favour of the one-Keplerian model given the combined HIRES and Lick data (case 2).}\label{combined_bf}
\begin{tabular}{lc}
\hline \hline
Estimate & $B$ \\
\hline
TPM & 2.0$\times10^{38} \pm  1.0\times10^{37}$ \\
DIC & 2.2$\times10^{38} \pm 1.9\times10^{37}$ \\
AIC & 5.7$\times10^{38}$ \\
OBMH & 2.0$\times10^{41} \pm 9.0\times10^{39}$ \\
HM & 1.4$\times10^{38} \pm 1.7\times10^{38}$ \\
\hline \hline
\end{tabular}
\end{table}

\subsection{Case 3. Partial HIRES data}

As a third test, we calculate the different estimates for marginal integral given only 20 epochs of HIRES data -- the first 20 epochs between 366 and 2602 JD-2450000 -- to see their relative performance when the number of parameters is comparable to the number of measurements. We find that the TPM converges to the marginal integral very accurately with $\lambda = 10^{-3}$ for both models and yields very reliable estimates for these integrals. It is again very close to the DIC estimate, making it reliable because of the Gaussianity of the posterior density for both models and the consequent reliability of the DIC estimate. It is not surprising that the AIC overestimates the Bayes factor and therefore also the posterior odds of the one-Keplerian model because of the low number of data. However, the OBMH overestimates it as well as was in fact found to be the case in test cases 1 and 2 as well.

\begin{table}
\center
\caption{Bayes factors in favour of the one-Keplerian model given the partial HIRES data (case 3).}\label{partial_bf}
\begin{tabular}{lc}
\hline \hline
Estimate & $B$ \\
\hline
TPM & 3.0$\times10^{5} \pm 5.1\times10^{4}$ \\
DIC & 2.8$\times10^{5} \pm 8.1\times10^{4}$ \\
AIC & 1.2$\times10^{9}$ \\
OBMH & 1.4$\times10^{6} \pm 3.3\times10^{5}$ \\
HM & 4.3$\times10^{3} \pm 8.1\times10^{3}$ \\
\hline \hline
\end{tabular}
\end{table}

\section{Artificial data: effect of prior choice}

We demonstrate further the properties of the TPM estimate by comparing its performance to more traditional integral estimation techniques. We generated four sets of artificial radial velocity data and determined the number of Keplerian signals using the TPM estimate and an estimate received using \emph{brute force} approach, i.e. direct numerical integration of the product of likelihood and prior over the parameter space. To demonstrate the conclusions in Section 3, we use an improper unit prior, i.e. $\pi_{1} = \pi(\theta) = 1$, and a broad prior of \citet{ford2007} with $P_{min} = P_{i}$, denoted as $\pi_{2}$, to show how they affect the conclusions that can be drawn from the same data.

The artificial data sets were generated by using 200 random epochs such that the first epoch was at $t=0$ and the $i$th one was selected randomly 1-10 days later within and interval of 7.2 hours, which simulates the fact that observations can only be made during the night. We generated the velocities by using a sinusoid with a period of 50 days and an amplitude of $K$ and added Gaussian random noise with zero mean and a variance of $1+ \sigma_{i}^{2}$, where $\sigma_{i}$ describes the standard deviation of the artificial Gaussian instrument noise. The values $\sigma_{i}$ were drawn from a uniform density between 0.3 and 0.6 for every simulated measurement. We generated sets S1, ..., S4 by using $K= 1.0, 0.8, 0.6, 0.5$, respectively.

We show the model comparison results of the four artificial data sets in Table \ref{artificial_comparisons}. This Table contains the Bayes factors in favour of the model with one Keplerian signal and against a model with no signals at all. We show the estimates calculated using a direct \emph{brute force} numerical integration (BF) for the two priors ($\pi_{1}$ and $\pi_{2}$) and the TPM estimate (that has approximately the same values for both priors, so we show only the results for $\pi_{1}$). These Bayes factors show, that the Bartlett's paradox clearly prevents the detection \citep[i.e. a Bayes factor in excess of 150; ][]{kass1995,tuomi2012} of the periodic signals in data set S3, whereas the TPM estimate, that does not fall victim to this paradox, yields a positive detection. The signal in the set S4 is too weak for detection.

\begin{table}
\center
\caption{Bayes factors in favour of model $\mathcal{M}_{1}$ for data sets S1, ..., S4 received using TPM estimate and the \emph{brute force} (BF) approach for two priors, $\pi_{1}$ and $\pi_{2}$.}\label{artificial_comparisons}
\begin{tabular}{lccc}
\hline \hline
Data & TPM & BF $\pi_{1}$ & BF $\pi_{2}$ \\
\hline
S1 & 5.0$\times 10^{15}$ & 3.7$\times 10^{13}$ & 3.1$\times 10^{8}$ \\
S2 & 5.3$\times 10^{9}$ & 5.6$\times 10^{7}$ & 4.7$\times 10^{2}$ \\
S3 & 1.2$\times 10^{3}$ & 71 & 6.0$\times 10^{-4}$ \\
S4 & 35 & 0.88 & 7.5$\times 10^{-6}$ \\
\hline \hline
\end{tabular}
\end{table}

It can be seen in Table \ref{artificial_comparisons} that the TPM estimate yields Bayes factors that support the existence of a signal in the data sets S1 - S3. In fact, the only data set where the signal could not be detected (S4), the Markov chains did not converge to a clear maximum in the period space either but several small maxima out of which none could be said to be significantly more probable than the others. In all the rest, the chains converged to a clear maximum corresponding to the periodic signals added to the artificial data sets.

It can also be seen how the broader prior ($\pi_{2}$) changes the Bayes factors when estimating the marginal integrals by direct numerical integration. Relative to the unit prior ($\pi_{1}$), the Bayes factors are roughly a factor of $10^{5}$ lower for $\pi_{2}$, and actually only provide a detection of the signal in data set S2 by exceeding the 150 threshold only just. This shows that the $\pi_{2}$ corresponds to \emph{a priori} model probabilities that are by a factor of $10^{5}$ more in favour of the model without Keplerian signals -- clearly an undesirable side-effect of the priors of \citet{ford2007}. Yet, the TPM estimate, and the corresponding Bayes factors, turned out to have roughly the same values for both priors as suspected because any constant terms in the prior do not affect the TPM estimate. Therefore, the TMP estimate enables the detection of weaker signals in the data than estimates that depend on constant coefficients in the prior density, and consequently, affect the prior probabilities of the models.

\section{Conclusions}

Calculating the marginal integral for model selection purposes is generally a challenging computational problem. While there are several good estimates for these integrals, they are usually only applicable under certain limiting assumptions about the nature of the posterior density, the amount of parameters in the statistical model, or the number of measurements available. Therefore, we have introduced a new method for estimating these integrals in practice. Given the availability of a sample from the posterior density of model parameters, our truncated posterior mixture estimate is a reasonably accurate one and very easily calculated in practice. We have only assumed that a statistically representative sample drawn from the posterior density exists when deriving our posterior mixture estimate (see Appendix). Therefore, it is applicable to any model comparison problems in astronomy and other fields of scientific inquiry and is not restricted to problems where the posterior has a certain shape and dimension.

The comparisons of different estimates given the radial velocities of HD 3651 revealed that the TPM yields estimates very close to the DIC estimate, which is known to be a reliable one in case of Gaussian posterior density. In fact, we chose the HD 3651 as an example star because of the planet orbiting it is known to have an eccentric orbit that enables the Gaussianity of the probability distributions of eccentricity and the two angular parameters of the Keplerian model, namely, longitude of pericentre and mean anomaly. However, the simple small-sample version of the AIC proved reasonably accurate as well when the number of measurements well exceeded the number of free parameters of the model (e.g. Table \ref{combined_bf}). We also note that the OBMH estimate of \citet{chib2001}, while converging rapidly, tends to yield somewhat biased results that exaggerate the posterior probability of the more complicated model, making it possibly -- at least in the test cases of the current work -- prone to detections of false positives.

In practice, the TMP can be used by calculating its value directly from the sample drawn from the posterior density of the model parameters. Selecting a suitable value for parameter $\lambda$ is then of essence when calculating its value in practice. In all the three different test cases studied in this article, a choice of $\lambda = 10^{-4}$ yielded estimates that converged rapidly for all the models in all the test cases and resulted in posterior probabilities that differed little from those calculated using the DIC estimate. When the difference between the two models was the smallest (case 3.), there was practically no bias in the TMP estimate with respect to the DIC. Also, when the posterior odds of the one-Keplerian model was the greatest (case 2.), the TPM, with $\lambda = 10^{-4}$, overestimated the posterior probability of the one-Keplerian model by a factor of 10, though, in that case, the Bayes factor used in model selection was already so heavily in favour of the one-Keplerian model that this overestimation is not significant in practice in terms of being able to select the best model.

Because of the possible biases caused by too large $\lambda$, it would then be convenient in practice to calculate the TPM estimate using few different values of parameter $\lambda$. With the sample from the posterior density available, this could be done with little computational cost. Then, it would be possible to use the lowest value for $\lambda$ that still converges to receive a trustworthy TPM estimate and correspondingly trustworthy model selection results in any model selection problem.

Finally, because any constant coefficients in the prior probability densities have an effect on the marginal integrals by corresponding to different prior weights for different models, we have shown how the TPM estimate deals with this problem. Effectively, it corresponds to setting the constant coefficients in the prior equal to unity, which makes the TPM estimate independent of the unit choice of the parameters.

\begin{acknowledgements}
M. Tuomi is supported by RoPACS (Rocky Planets Around Cools Stars), a Marie Curie Initial Training Network funded by the European Commission's Seventh Framework Programme. The authors would like to acknowledge P. Gregory for fruitful comments that resulted in significant improvements of the article.
\end{acknowledgements}

\appendix

\section{Marginal integrals from importance sampling}

In the context of Bayesian model selection, the marginal integral needed to assess the relative probabilities of different model is
\begin{equation}\label{marginal_integral}
  P(m | \mathcal{M}) = \int_{\theta \in \Theta} l(m | \theta, \mathcal{M}) \pi(\theta | \mathcal{M}) d \theta ,
\end{equation}
where $\mathcal{M}$ is a model with parameter vector $\theta$ constructed to model the measurements $m$ using the likelihood function $l$. Function $\pi(\theta | \mathcal{M})$ is the prior probability density of the model parameters. This quantity is essential in calculating the posterior probabilities of different models in Eq. (\ref{model_probability}).

Importance sampling can be used to receive estimates for the integral in Eq. (\ref{marginal_integral}). Choosing functions $g$ and $w$ such that $\pi(\theta) = w(\theta) g(\theta)$ and dropping the model from the notation, the marginal integral can be written using the expectation with respect to density $g$ as
\begin{equation}\label{expectation}
  \mathbb{E}_{g}\big[ w(\theta) l(m | \theta) \big] = \int_{\theta \in \Theta} g(\theta) w(\theta) l(m | \theta) d \theta = P(m).
\end{equation}
where $g(\theta)$ is usually called the importance sampling function. Now, the idea of importance sampling is that if we draw a sample of $N$ values from $g$ and denote $\theta_{i} \sim g(\theta)$ for all $i = 1, ..., N$, we can calculate a simple estimate for the marginal integral as \citep[e.g.][]{kass1995}
\begin{equation}\label{marginal_estimate}
  \hat{P} = \Bigg[ \sum_{i=1}^{N} \frac{\pi(\theta_{i}) l(m | \theta_{i})}{g(\theta_{i})} \Bigg] \Bigg[ \sum_{i=1}^{N} \frac{\pi(\theta_{i})}{g(\theta_{i})} \Bigg]^{-1} .
\end{equation}
All there remains is to choose $g$ such that it is easy to draw a sample from it and that the estimate in Eq. (\ref{marginal_estimate}) converges rapidly to the marginal integral.

Some simple choices of $g$ would be the prior density or the posterior density. In these cases, the resulting estimates would be called the mean estimate and the harmonic mean estimate, respectively \citep{newton1994,kass1995}. We denote these estimates as $\hat{P}_{M}$ and $\hat{P}_{HM}$ and write
\begin{equation}\label{M_estimate}
  \hat{P}_{M} = \frac{1}{N} \sum_{i=1}^{N} l(m | \theta_{i})
\end{equation}
and
\begin{equation}\label{HM_estimate}
  \hat{P}_{HM} = N \Bigg[ \sum_{i=1}^{N} \frac{1}{l(m | \theta_{i})} \Bigg]^{-1} .
\end{equation}

Though easily computed in practice, these estimates have some undesirable properties. For instance, the mean estimate requires drawing a sample from the prior density and computation of the corresponding likelihoods. However, because the prior contains less information and is therefore much broader density than the posterior, most of the values in this sample correspond to very low likelihoods and the convergence of this estimate is generally slow. The resulting value is also dominated by few high likelihoods, which can make it too biased to be useful in applications, except in very simple cases.

Also, while converging to the desired value, the harmonic mean estimate does it extremely slowly in practice \citep{kass1995} and its usage cannot be recommended. In applications, this estimate doesn't generally converge to the marginal integral within the limited sample available from the posterior. The reason is that occasional small values of $l(m | \theta_{i})$ have a large impact on the sum which makes its convergence extremely slow. For these reasons, better estimates are needed to approximate the marginal integrals in model selection problems.

\subsection{The posterior mixture estimate}

To construct a better estimate for the marginal integral, we start by assuming that a statistically representative sample has been drawn from the posterior density using some posterior sampling algorithm. Therefore, we have a collection of $N$ vectors $\theta_{i} \sim \pi(\theta | m)$, for all $i = 1, ..., N$. These values form a Markovian chain with $N$ members. Selecting integer $h > 0$, the value of the posterior $\pi(\theta_{i-h} | m)$ is available if the value corresponding to $\theta_{i}$ is available given $i > h > 0$. Here we can denote $\pi_{i} = \pi(\theta_{i} | m)$ and see that if $\theta_{i}$ is a random vector then $\pi_{i}$ is some random number corresponding to the value of the posterior at $\theta_{i}$. Using the notation similarly for $g_{i}$, and setting $\lambda \in [0,1]$, we can set
\begin{equation}\label{definition_g}
  g_{i} = (1 - \lambda) \pi_{i} + \lambda \pi_{i-h} .
\end{equation}

Now, if $\lambda$ is a small number, it follows that $g_{i} \approx \pi_{i}$ -- the importance sampling function $g$ is close to the posterior but not exactly equal. We call it a truncated posterior mixture (TPM) function. The sample from the posterior is close to a sample from $g$ -- a desired property because a sample from the posterior can be calculated rather readily with posterior sampling algorithms \citep[e.g.][]{metropolis1953,hastings1970,haario2001}. The estimate in Eq. (\ref{marginal_estimate}) can now be calculated. We denote $l_{i} = l(m | \theta_{i})$ and $p_{i}= \pi(\theta_{i})$ and write the resulting posterior mixture estimate as
\begin{eqnarray}\label{TPM_estimate}
 && \hat{P}_{TPM} = \Bigg[ \sum_{i=1}^{N} \frac{l_{i}p_{i}}{(1-\lambda) l_{i}p_{i} + \lambda l_{i-h}p_{i-h}} \Bigg] \nonumber\\
 && \times \Bigg[ \sum_{i=1}^{N} \frac{p_{i}}{(1-\lambda) l_{i}p_{i} + \lambda l_{i-h}p_{i-h}} \Bigg]^{-1} .
\end{eqnarray}

If the Markov chain has good mixing properties such that the value $\theta_{i}$ has already become independent of $\theta_{i-h}$, the likelihoods of these values are also independent. When comparing this estimate with $\hat{P}_{HM}$ in Eq. (\ref{HM_estimate}), it can be seen that occasional small values of $l_{i}$ do not have such a large effect on the sum in the denominator because it is unlikely that the corresponding value of $l_{i-h}$ is also small at the same time.



\begin{thebibliography}{100}\small
\bibitem[\protect\astroncite{Akaike}{1973}]{akaike1973} Akaike, H. 1973. pp. 267 in Petrov, B. N. \& Csaki, F. (eds.). Second International Symposium on Information Theory. Akad\'emiai Kiad\'o. 1
\bibitem[\protect\astroncite{Barge et al.}{2007}]{barge2007} Barge, P., Baglin, A., Auvergne, M., et al. 2007, A\&A, 482, L17
\bibitem[\protect\astroncite{Bartlett}{1957}]{bartlett1957} Bartlett, M. S. 1957, Biometrika, 44, 533 
\bibitem[\protect\astroncite{Bean et al.}{2010}]{bean2010} Bean, J. L., Seifahrt, A., Hartman, H., et al. 2010, The ApJ, 711, L19
\bibitem[\protect\astroncite{Benedict et al.}{2002}]{benedict2002} Benedict, G. F., McArthur, B. E., Forveille, T., et al. 2002, ApJ, 581, L115
\bibitem[\protect\astroncite{Burnham \& Anderson}{2002}]{burnham2002} Burnham, K. P. \& Anderson D. R. 2002, Model selection and multimodel inference: A practical information-theoretic approach, Springer-Verlag
\bibitem[\protect\astroncite{Butler et al.}{2006}]{butler2006} Butler, R. P., Wright, J. T., Marcy, G. W., et al. 2006, ApJ, 646, 505
\bibitem[\protect\astroncite{Chib \& Jeliazkov}{2001}]{chib2001} Chib S. \& Jeliazkov I. 2001, J. Am. Stat. Ass., 96, 270
\bibitem[\protect\astroncite{Clyde et al.}{2007}]{clyde2007} Clyde, M. A., Berger, J. O., Bullard, F., et al. 2007, Statistical Challenges in Modern Astronomy IV, Babu, G. J. \& Feigelson, E. D. (eds.), ASP Conf. Ser., 371, 224
\bibitem[\protect\astroncite{Collier Cameron et al.}{2007}]{colliercameron2007} Collier Cameron, A., Bouchy, F., H\'ebrard, G., et al. 2007, MNRAS, 375, 951
\bibitem[\protect\astroncite{Faedi et al.}{2011}]{faedi2011} Faedi, F., Barros, S. C. C., Anderson, D. R., et al. 2011, A\&A, 531, A40
\bibitem[\protect\astroncite{Feroz et al.}{2011}]{feroz2011} Feroz, F., Balan, S. T., \& Hobson, M. P. 2011, MNRAS, 415, 3462
\bibitem[\protect\astroncite{Fischer et al.}{2003}]{fischer2003} Fischer, D. A., Butler, R. P., Marcy, G. W., et al. 2003, ApJ, 590, 1081
\bibitem[\protect\astroncite{Ford \& Gregory}{2007}]{ford2007} Ford, E. B. \& Gregory, P. C. 2007, Statistical Challenges in Modern Astronomy IV, Babu, G. J. \& Feigelson, E. D. (eds.), ASP Conf. Ser., 371, 189
\bibitem[\protect\astroncite{Forveille et al.}{2011}]{forveille2011} Forveille, T., Bonfils, X., Delfosse, X., et al. 2011, A\&A, submitted (arXiv:1109.2505 [astro-ph.EP])
\bibitem[\protect\astroncite{Gregory}{2005}]{gregory2005} Gregory, P. C. 2005, ApJ, 631, 1198
\bibitem[\protect\astroncite{Gregory}{2007a}]{gregory2007a} Gregory, P. C. 2007a, MNRAS, 381, 1607
\bibitem[\protect\astroncite{Gregory}{2007b}]{gregory2007b} Gregory, P. C. 2007b, MNRAS, 374, 1321
\bibitem[\protect\astroncite{Gregory}{2011}]{gregory2011} Gregory, P. C. 2011, MNRAS, 415, 2523
\bibitem[\protect\astroncite{Haario et al.}{2001}]{haario2001} Haario, H., Saksman, E., \& Tamminen, J. 2001, Bernoulli, 7, 223
\bibitem[\protect\astroncite{Hastings}{1970}]{hastings1970} Hastings, W. 1970, Biometrika 57, 97
\bibitem[\protect\astroncite{H\'ebrard et al.}{2011}]{hebrard2011} H\'ebrard, G., Evans, T. M., Alonso, R., et al. 2011. A\&A, 533, A130
\bibitem[\protect\astroncite{Holman \& Murray}{2005}]{holman2005} Holman, M. J. \& Murray, N. W. 2005, Science, 307, 1288
\bibitem[\protect\astroncite{Jeffreys}{1961}]{jeffreys1961} Jeffreys, H., 1961, Theory of Probability, Oxford University Press
\bibitem[\protect\astroncite{Jones et al.}{2002}]{jones2002} Jones, H. R. A., Butler, R. P., Marcy, G. W., et al. 2002, MNRAS, 337, 1170
\bibitem[\protect\astroncite{Kass \& Raftery}{1995}]{kass1995} Kass, R. E. \& Raftery, A. E. 1995, J. Am. Stat. Ass., 430, 773
\bibitem[\protect\astroncite{Kullback \& Leibler}{1951}]{kullback1951} Kullback, S. \& Leibler, R. A. 1951, Ann. Math. Stat., 22, 76
\bibitem[\protect\astroncite{Lovis et al.}{2011}]{lovis2011} Lovis, C., S\'egransan, D., Mayor, M., et al. 2011, A\&A, 528, A112
\bibitem[\protect\astroncite{Mayor et al.}{2003}]{mayor2003} Mayor, M., Pepe, F., Queloz, D., et al. 2003, Messenger, 114, 20
\bibitem[\protect\astroncite{Metropolis et al.}{1953}]{metropolis1953} Metropolis, N., Rosenbluth, A. W., Rosenbluth, M. N., et al. 1953, J. Chem. Phys., 21, 1087
\bibitem[\protect\astroncite{Newton \& Raftery}{1994}]{newton1994} Newton, M. A. \& Raftery, A. E. 1994, J. Roy. Stat. Soc. B, 56, 3
\bibitem[\protect\astroncite{Pravdo \& Shaklan}{2009}]{pravdo2009} Pravdo, S. H. \& Shaklan, S. B. 2009, ApJ, 700, 623
\bibitem[\protect\astroncite{Spiegelhalter et al.}{2002}]{spiegelhalter2002} Spiegelhalter, D. J., Best, N. G., Carlin, B. P., \& van der Linde, A. 2002, JRSS B, 64, 583
\bibitem[\protect\astroncite{Tinney et al.}{2001}]{tinney2001} Tinney, C. G., Butler, R. P., Marcy, G. W., et al. 2001, ApJ, 551, 507
\bibitem[\protect\astroncite{Tuomi \& Kotiranta}{2009}]{tuomi2009} Tuomi, M. \& Kotiranta, S. 2009, A\&A, 496, L13
\bibitem[\protect\astroncite{Tuomi}{2011}]{tuomi2011} Tuomi, M. 2011, A\&A, 528, L5
\bibitem[\protect\astroncite{Tuomi}{2012}]{tuomi2012} Tuomi, M. 2012, A\&A, in press (arXiv:1204.1254 [astro-ph.EP])
\bibitem[\protect\astroncite{Tuomi et al.}{2011}]{tuomi2011b} Tuomi, M., Pinfield, D., \& Jones, H. R. A. 2011, A\&A, 532, A116
\bibitem[\protect\astroncite{Vogt et al.}{1994}]{vogt1994} Vogt, S. S., Allen, S. L., Bigelow, B. C., et al. 1994, SPIE Instrumentation in Astronomy VIII, Crawford, D. L., Craine, E. R. (eds.), 2198, p. 362
\bibitem[\protect\astroncite{Vogt et al.}{2010}]{vogt2010} Vogt, S. S., Butler, R. P., Rivera, E. J., et al. 2010, ApJ, 723, 954
\end{thebibliography}
\end{document}